\documentclass[aps,prc,twocolumn,showpacs,superscriptaddress,groupedaddress,floatfix]{revtex4-1}
\usepackage[english]{babel}
\usepackage[T1]{fontenc}
\usepackage{fontenc}
\usepackage{graphicx}
\usepackage{subfigure}
\usepackage{bm,amssymb,amsmath}
\usepackage{booktabs}
\usepackage{multirow}
\usepackage[titletoc]{appendix}
\usepackage[colorlinks=true,linkcolor=blue,citecolor=blue,urlcolor=blue]{hyperref}
\DeclareMathOperator{\e}{e}
\newcommand{\goodgap}{%
	\hspace{\subfigtopskip}%
	\hspace{\subfigbottomskip}}

\begin{document}

\title{Accurate determination of the interaction between $\Lambda$ hyperons and 
nucleons from auxiliary field diffusion Monte Carlo calculations}

\author{D. Lonardoni}
\email{lonardoni@science.unitn.it}
\affiliation{Physics Department, University of Trento, via Sommarive, 14 I-38123 Trento, Italy}
\affiliation{INFN - TIFPA, Trento Institute for Fundamental Physics and Applications, Trento, Italy}

\author{S. Gandolfi}
\affiliation{Theoretical Division, Los Alamos National Laboratory, Los Alamos, 87545 US-NM}

\author{F. Pederiva}
\email{pederiva@science.unitn.it}
\affiliation{Physics Department, University of Trento, via Sommarive, 14 I-38123 Trento, Italy}
\affiliation{INFN - TIFPA, Trento Institute for Fundamental Physics and Applications, Trento, Italy}

\begin{abstract}
	\begin{description} 
		\item[Background] An accurate assessment of the hyperon-nucleon interaction is
		of great interest in view of recent observations of very massive neutron stars. 
		The challenge is to build a realistic interaction that can be used over a wide range of 
		masses and in infinite matter 
		starting from the available experimental data on the binding energy of light hypernuclei.
		To this end, accurate calculations of the hyperon binding energy in a hypernucleus are necessary.
		\item[Purpose] We present a quantum Monte Carlo study of $\Lambda$ and $\Lambda\Lambda$
		hypernuclei up to $A=91$. We investigate the contribution of two- and three-body
		$\Lambda$-nucleon forces to the $\Lambda$ binding energy.
		\item[Method] Ground state energies are computed solving the Schr\"odinger equation for
		non-relativistic baryons by means of the auxiliary field diffusion Monte
		Carlo algorithm extended to the hypernuclear sector. 
		\item[Results] We show that a simple adjustment of the
		parameters of the $\Lambda NN$ three-body force yields a very good
		agreement with available experimental data over a wide range of hypernuclear masses.
		In some cases no experiments have been performed yet, and we give new predictions.
		\item[Conclusions] The newly fitted $\Lambda NN$ force properly describes the physics of 
		medium-heavy $\Lambda$~hypernuclei, correctly reproducing the saturation property of 
		the hyperon separation energy.
	\end{description}
\end{abstract}


\maketitle

\section{Introduction}
\label{sec:intro}

The problem of determining a realistic interaction among hyperons and nucleons 
capable of reconciling the terrestrial measurements on hypernuclei and the recent 
observations of very massive neutron stars is still by and large unsolved. 
The amount of data available for nucleon-nucleon scattering~\cite{Stoks:1993}
is enough to build satisfactory models of nuclear forces, either purely 
phenomenological or built on the basis of an effective field theory
\cite{Stoks:1994,Wiringa:1995,Machleidt:1996,Wiringa:2002,Epelbaum:2005,
Ekstrom:2013,Gezerlis:2013}. In the hyperon-nucleon sector, much less data are 
available~\cite{deSwart:1971,Kadyk:1971,Ahn:2005}, and almost nothing is known 
about the hyperon-hyperon interaction. The main reasons of this lack of 
information are the instability of hyperons in the vacuum and the impossibility 
of collecting hyperon-neutron and hyperon-hyperon scattering data. This implies that 
interaction models must be fitted mostly to binding energies (and possibly excitations) 
of hypernuclei.

Besides the very old emulsion data~\cite{Juric:1973,Cantwell:1974},
several measurements of hypernuclear energies have become available
in the last few years~\cite{Pile:1991,Hasegawa:1996,Takahashi:2001,Yuan:2006,Cusanno:2009,
Agnello:2010,Nakazawa:2010,Agnello:2012_H6L,Nakamura:2013,Ahn:2013,Feliciello:2013},
both for single and double $\Lambda$~hypernuclei. These can be used to validate 
or to constrain the hyperon-nucleon interactions within the framework of many-body systems.
The ultimate goal is then to constrain these forces by reproducing
the experimental energies of hypernuclei from light systems
made of few particles up to heavier systems.

In previous work it was shown that the inclusion of a  $\Lambda NN$ interaction 
gives a very important repulsive contribution towards a realistic description of the
saturation property of the $\Lambda$~separation energy 
in medium-heavy hypernuclei~\cite{Lonardoni:2013_PRC(R)}. 
In this paper, we focus on single and double $\Lambda$~hypernuclei
to study more in detail the role of $\Lambda N$ and $\Lambda NN$ interactions.
Ground state properties of hypernuclei are here computed by means of Quantum Monte 
Carlo (QMC) methods, and in particular by the auxiliary field diffusion Monte Carlo 
(AFDMC) algorithm. These methods have been shown to be very accurate in solving the 
many-body problem fully non-perturbatively even when the system is dominated by 
very strong correlations that cannot be neglected~\cite{Pieper:2008}. 
This is the case for nuclear systems.

The paper is organized as follows. In Sec.~\ref{sec:hamiltonians},
we introduce the Hamiltonians involved in the description of single
and double $\Lambda$~hypernuclei. Section~\ref{sec:method} gives
an overview of the auxiliary field diffusion Monte Carlo method
with particular attention to its application to hypernuclear systems
(Sec.~\ref{subsec:AFDMChyp}). Next, in Sec.~\ref{sec:results} we report
the results for single $\Lambda$~hypernuclei (Sec.~\ref{subsec:singleL})
and for double $\Lambda$~hypernuclei (Sec.~\ref{subsec:doubleL}).
Finally, in Sec.~\ref{sec:conclusions} the conclusions of our work.

\section{Hamiltonian}
\label{sec:hamiltonians}

We describe nuclei and $\Lambda$~hypernuclei with a non-relativistic
Hamiltonian that includes two- and three-body forces,
\begin{align}
H_{\rm nuc}&=T_N+V_{NN}=\sum_{i}\frac{p_i^2}{2m_N}+\sum_{i<j}v_{ij}\;,\\[0.7em]
H_{\rm hyp}&=H_{\rm nuc}+T_\Lambda+V_{\Lambda N}+V_{\Lambda NN}+V_{\Lambda\Lambda}\nonumber\\[0.5em]
&=H_{\rm nuc}+\sum_{\lambda}\frac{p_\lambda^2}{2m_\Lambda}+\sum_{\lambda i}v_{\lambda i}
+\!\!\sum_{\lambda,i<j}v_{\lambda ij}+\!\sum_{\lambda<\mu}v_{\lambda\mu}\;,
\end{align}
where $A$ is the total number of baryons $A=\mathcal
N_N+\mathcal N_\Lambda$, latin indices $i,j=1,\ldots,\mathcal N_N$ label
nucleons, and greek symbols $\lambda,\mu=1,\ldots,\mathcal N_\Lambda$
are used for $\Lambda$~particles. The nuclear potential is limited to a
two-body interaction, while in the strange sector we adopt explicit
$\Lambda N$ and $\Lambda NN$ interactions. In the case of double
$\Lambda$~hypernuclei, a $\Lambda\Lambda$ force is also involved.

\subsection{Nucleon-Nucleon interaction}
\label{subsec:NN}

The interaction between nucleons is described via the Argonne V4'
and V6' two body-potentials~\cite{Wiringa:2002}, that are simplified
versions of the more sophisticated Argonne V18 potential~\cite{Wiringa:1995}
obtained with a re-projection of the interaction to preserve the
phase shifts of lower partial waves. The Argonne potential between
two nucleons $i$ and $j$ is written in coordinate space as a sum of
operators 
\begin{align}
	v_{ij}=\sum_{p=1}^n v_p(r_{ij})\mathcal O_{ij}^{\,p}\;,
\end{align} where $n$ is the number of operators, which depends on
the potential, $v_p(r_{ij})$ are radial functions, and $r_{ij}$ is the
interparticle distance. The six operators included in the Argonne
V6' potential mainly come from the one-pion exchange (OPE) between nucleons and they
read 
\begin{align}
	\mathcal O_{ij}^{\,p=1,6}=\left(1,{\bm\sigma_i}\cdot{\bm\sigma_j},S_{ij}\right)
	\otimes\left(1,{\bm\tau_i}\cdot{\bm\tau_j}\right)\;, \label{eq:op}
\end{align} 
where $S_{ij}$ is the usual tensor operator 
\begin{align}
	S_{ij}=3\left(\bm\sigma_i\cdot \hat{\bm r}_{ij}\right)
	\left(\bm\sigma_j\cdot \hat{\bm r}_{ij}\right)
	-\bm\sigma_i\cdot\bm\sigma_j\;. \label{eq:S_ij}
\end{align}
The AV4' force does not include the tensor terms $p=5,6$.

It is important to note that the above nuclear potentials do
not provide the same accuracy as AV18 in fitting $NN$ scattering
data in all partial waves. In addition, three-body $NNN$ forces
are purposely disregarded for technical reasons related to the
AFDMC algorithm used. As reported
in Refs.~\cite{Lonardoni:2013_HYP2012,Lonardoni:2013_PRC(R)}, these restrictions on
the nuclear potentials do not affect the main result of this work, namely the
calculation of the $\Lambda$~separation energy (the difference between the
binding energy of a nucleus and the corresponding $\Lambda$~hypernucleus),
which is not significantly dependent on the specific choice of the
nucleon Hamiltonian.

\subsection{Hyperon-Nucleon interaction}
\label{subsec:YN}

To describe the interaction between the $\Lambda$~particle and
the nucleons, we adopt a class of Argonne-like interactions
that have been developed starting from the 1980's by Bodmer,
Usmani, and Carlson on the grounds of Quantum Monte Carlo methods
and have been mostly used in Variational Monte Carlo 
calculations~\cite{Bodmer:1984,Bodmer:1985,Bodmer:1988,Usmani:1995,Shoeb:1998,
Shoeb:1999,Usmani:1999,Sinha:2002,Usmani:2003,Usmani:2004,Shoeb:2004,Usmani:2006,
Usmani:2006_He6LL,Usmani:2008}.
The interaction is written in coordinate space and it includes two-
and three-body hyperon-nucleon components with a hard-core repulsion
between baryons and a charge symmetry breaking term.
\begin{figure*}[!t]
	\centering
		\subfigure[\label{fig:LN_2pi}]{\includegraphics[height=3.2cm]{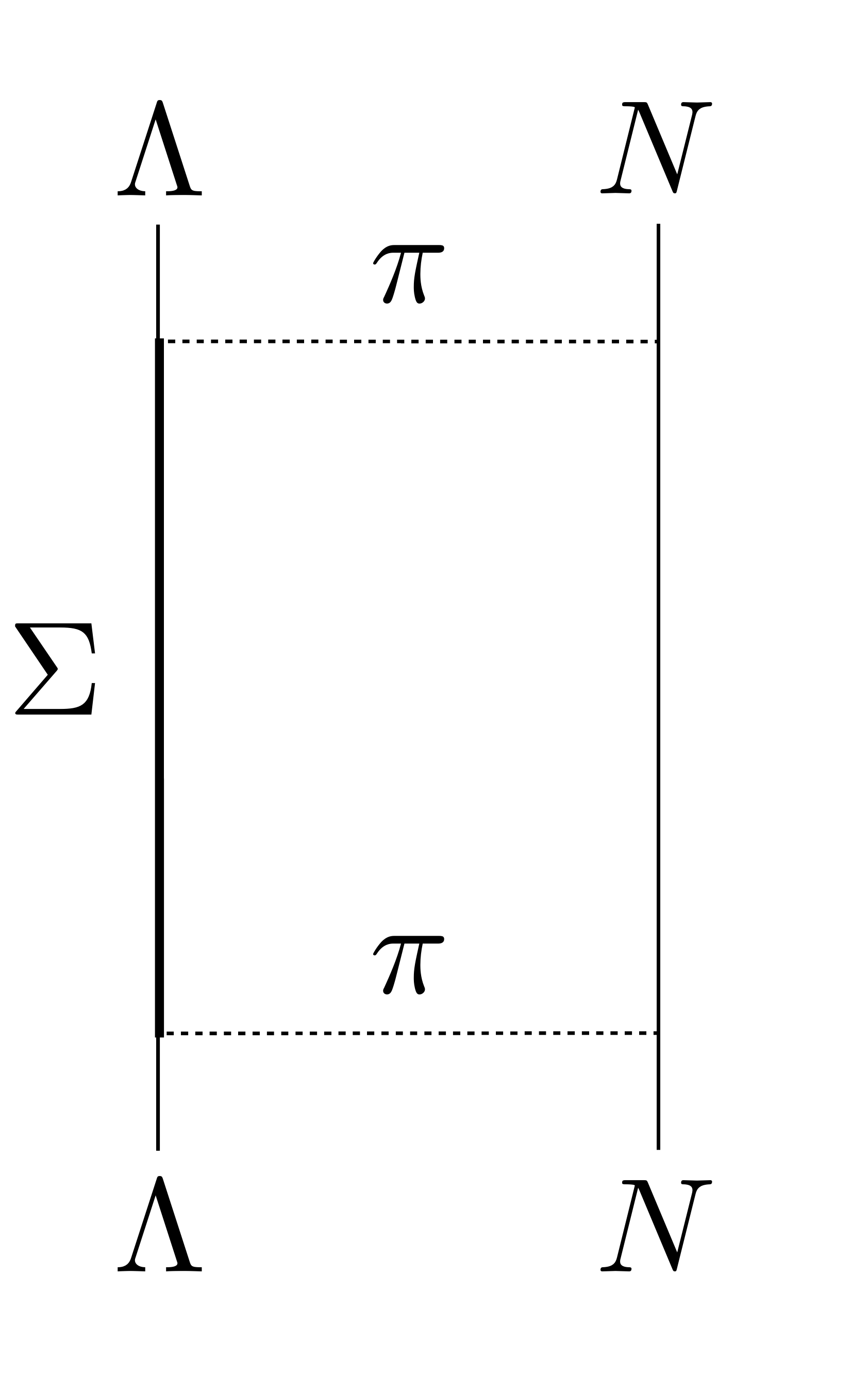}}
		\goodgap\goodgap\goodgap
		\subfigure[\label{fig:LN_K}]{\includegraphics[height=3.2cm]{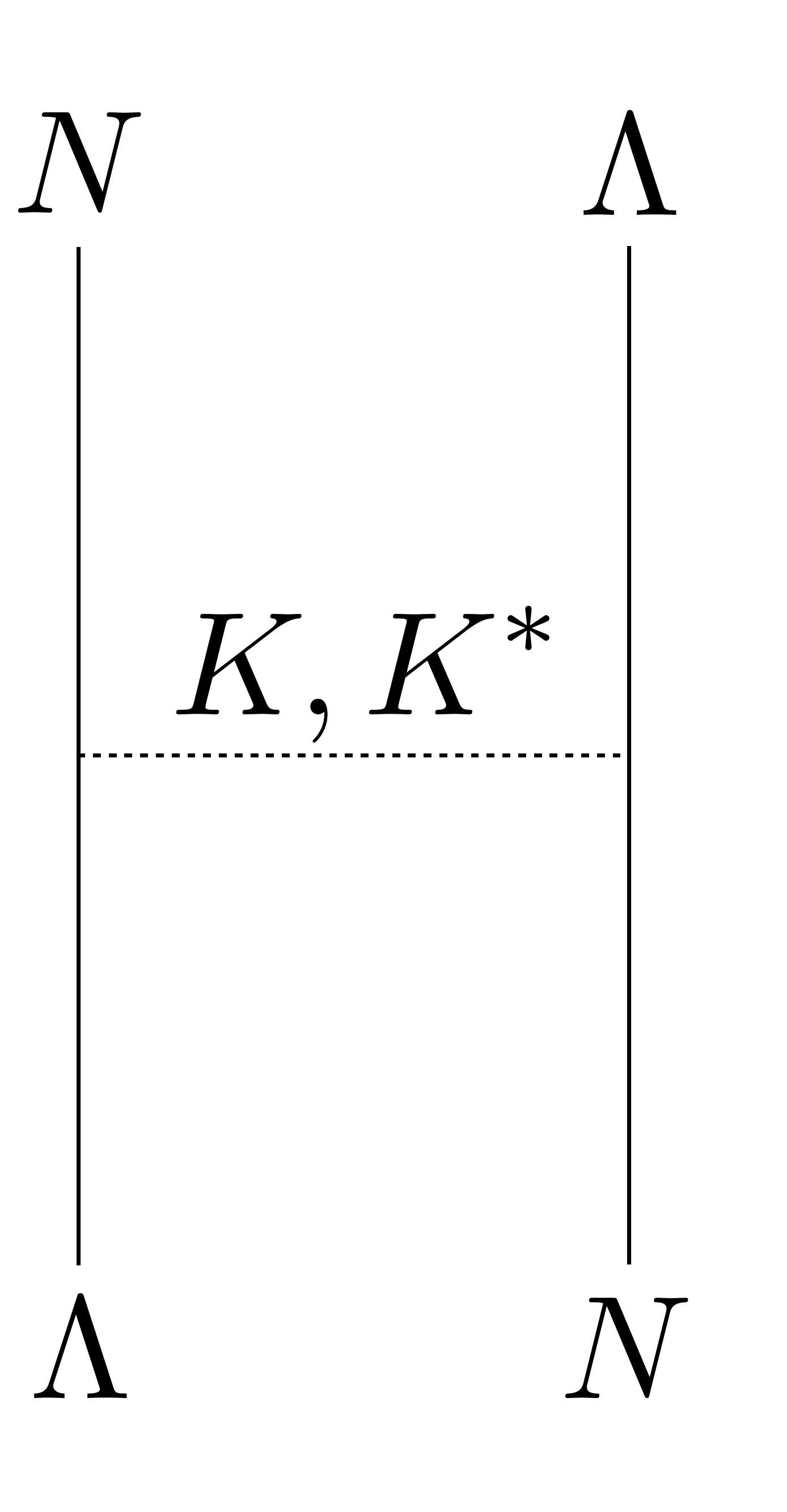}}
		\goodgap\goodgap\goodgap
		\subfigure[\label{fig:LNN_pw}]{\includegraphics[height=3.2cm]{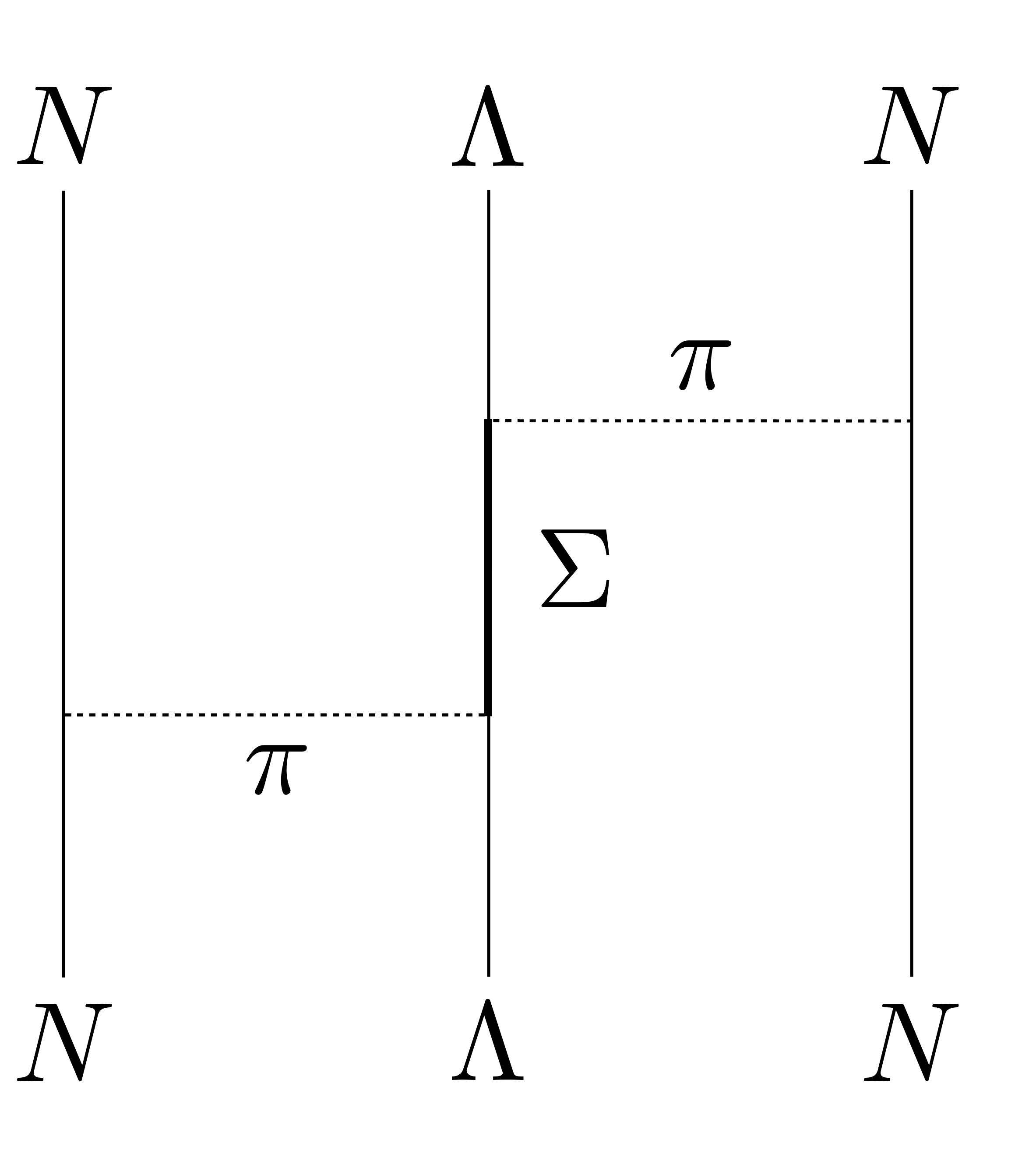}}
		\goodgap\goodgap\goodgap
		\subfigure[\label{fig:LNN_sw}]{\includegraphics[height=3.2cm]{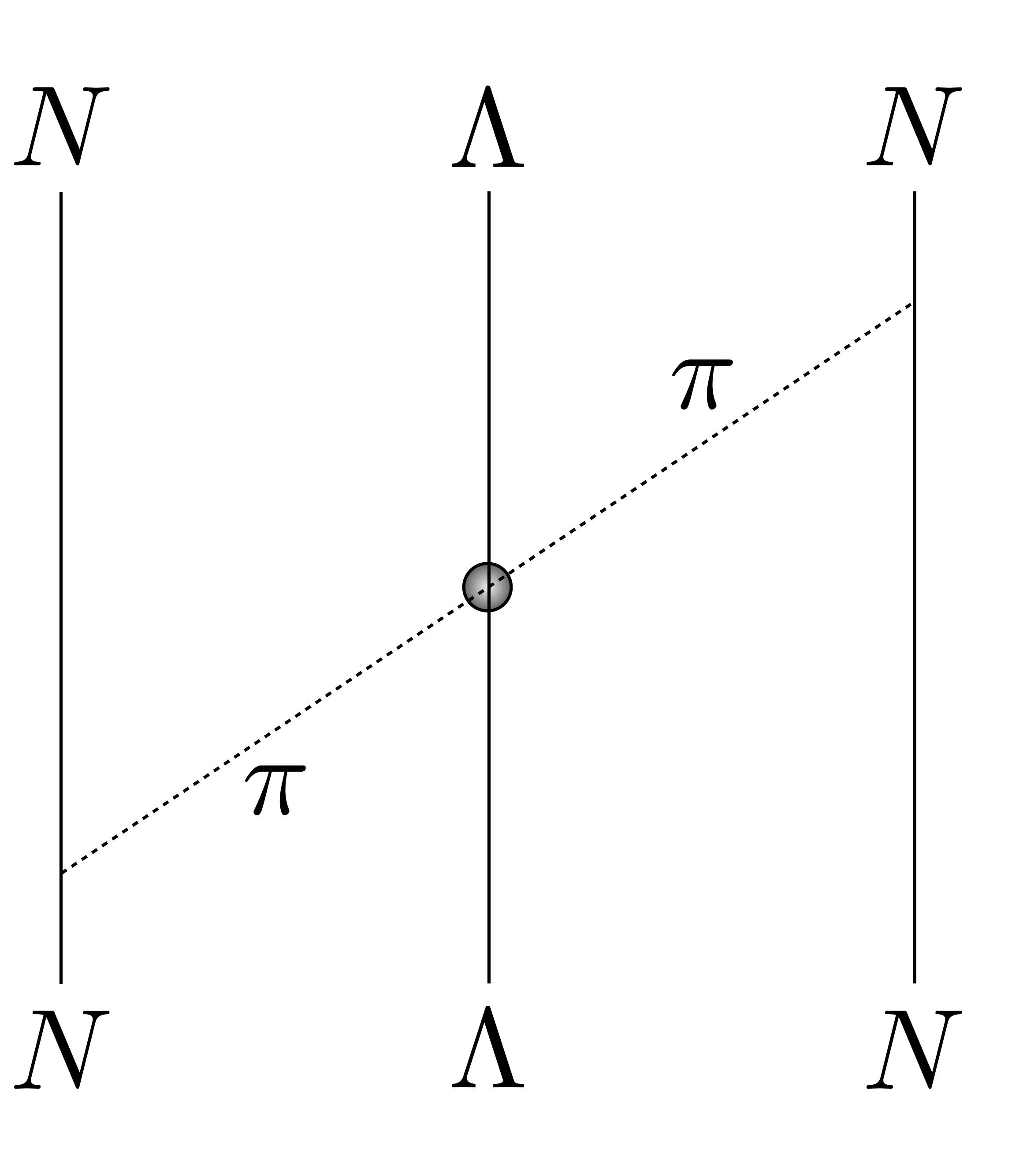}}
		\goodgap\goodgap\goodgap
		\subfigure[\label{fig:LNN_d}]{\includegraphics[height=3.2cm]{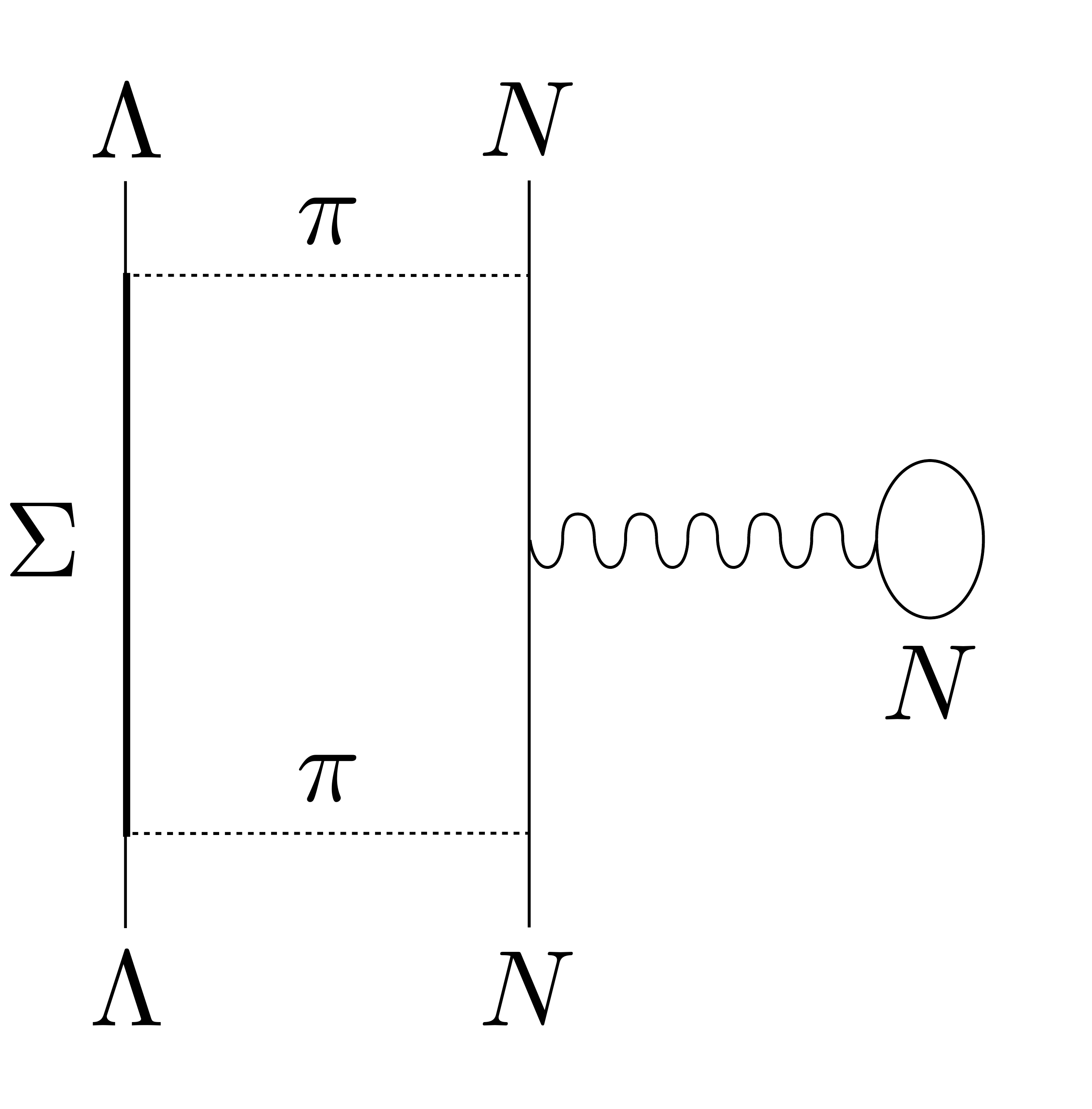}}
		\caption[]{Meson exchange processes between nucleons and hyperons. 
			\ref{fig:LN_2pi} and \ref{fig:LN_K} represent the $\Lambda N$ channels.
			\ref{fig:LNN_pw}-\ref{fig:LNN_d} are the three-body $\Lambda NN$ channels 
			included in the potential by Usmani~\emph{et al.}. 
			See~\cite{Usmani:2008} and references therein.}
		\label{fig:diagrams} 
\end{figure*}

\emph{$\Lambda N$ charge symmetric potential.}\label{subsubsec:LN}
Since the $\Lambda$~particle has isospin $I=0$, there is no OPE term, 
the strong $\Lambda\pi\Lambda$ vertex being
forbidden due to isospin conservation.  The $\Lambda$ hyperon can
exchange a pion only via a $\Lambda\pi\Sigma$ vertex.  The lowest-order
$\Lambda N$ coupling must therefore involve the exchange of two pions,
with the formation of a virtual $\Sigma$ hyperon, as illustrated in
Fig.~\ref{fig:LN_2pi}. The 2$\pi$-exchange interaction is of intermediate
range with respect to the long-range part of $NN$ force.  One-meson-exchange 
processes can only occur through the exchange of a $K,K^*$
kaon pair, that contributes in exchanging strangeness between the two
baryons, as shown in Fig.~\ref{fig:LN_K}. The $K,K^*$ exchange potential
is short-range and it is expected to be quite weak because the $K$
and $K^*$ tensor contributions have opposite sign~\cite{Shinmura:1984}.
The short-range contributions are included, as in the Argonne $NN$
interaction, by means of a Wood-Saxon repulsive potential
\begin{align}
	v_c(r)=W_c\Bigl(1+\e^{\frac{r-\bar r}{a}}\Bigr)^{-1}\;.
\end{align}
The $\Lambda N$ interaction has therefore been modeled with an Urbana-type 
potential~\cite{Lagaris:1981}, consistent with the available $\Lambda p$
scattering data, with an explicit space exchange term
\begin{align}
	\!\!\!v_{\lambda i}=v_{0}(r_{\lambda i})(1-\varepsilon+\varepsilon\,\mathcal P_x)
	+\frac{1}{4}v_\sigma T^2_\pi(r_{\lambda i})\,{\bm\sigma}_\lambda\cdot{\bm\sigma}_i \;,
	\label{eq:V_LN}
\end{align}
where $\mathcal P_x$ is the $\Lambda N$ exchange operator and
$v_0(r)=v_c(r)-\bar v\,T_{\pi}^{2}(r)$ is a central term. The terms
$\bar v=(v_s+3v_t)/4$ and $v_\sigma=v_s-v_t$ are the spin-average and
spin-dependent strengths, where $v_s$ and $v_t$ denote singlet- and
triplet-state strengths, respectively.  Note that both the spin-dependent
and the central radial terms contain the usual regularized OPE tensor
operator $T_\pi(r)$
\begin{align}
	T_{\pi}(r)=\left[1+ \frac{3}{\mu_\pi r}+ \frac{3}{(\mu_\pi r)^2} \right]
	\frac{\e^{-\mu_\pi r}}{\mu_\pi r}\Bigl(1-\e^{-cr^2}\Bigr)^2\;,\label{eq:T_pi}
\end{align}
where $\mu_\pi$ is the pion reduced mass
\begin{align}
	\mu_\pi=\frac{1}{\hbar}\frac{m_{\pi^0}+2\,m_{\pi^\pm}}{3}\quad\quad\frac{1}{\mu_\pi}\simeq 1.4~\text{fm} \;.
\end{align}
All the parameters defining the $\Lambda N$ potential can be found
in Table~\ref{tab:parLN+LNN}. For more details see for example
Ref.~\cite{Usmani:2008}.

\emph{$\Lambda N$ charge symmetry breaking potential.}\label{subsubsec:CSB}
The $\Lambda$-nucleon interaction should distinguish between the nucleon
isospin channels $\Lambda p$ and $\Lambda n$.  This is required by the
experimental data, in particular the $^4_\Lambda$H and $^4_\Lambda$He
ground- and excited-state energies~\cite{Juric:1973}, that have been
reproduced in Ref.~\cite{Bodmer:1985} by means of a phenomenological
spin-dependent, charge-symmetry breaking (CSB) potential. It was found
that the CSB contribution is effectively spin independent.  Following
Ref.~\cite{Usmani:1999}, we can express the CSB $\Lambda N$ interaction as
\begin{align}
	v_{\lambda i}^{CSB}=
	C_\tau\,T_\pi^2\left(r_{\lambda i}\right)\tau_i^z\;,
	\label{eq:V_CSB}
\end{align}
where $C_\tau$ was found by the analysis
of the $A=4$ mirror $\Lambda$~hypernuclei and it is listed in
Table~\ref{tab:parLN+LNN}. $C_\tau$ being negative,
the $\Lambda p$ channel becomes attractive while the $\Lambda n$
channel is repulsive. The contribution of CSB is expected to be very
small in symmetric hypernuclei (if Coulomb is neglected) but could have
a significant effect in hypernuclei with a neutron (or proton) excess.

\emph{$\Lambda NN$ potential.}\label{subsubsec:LNN}
In contrast to the nucleon-nucleon force, the lowest order 
$\Lambda$-nucleon interaction involves the exchange of two pions.
At the same 2$\pi$-exchange order, there are diagrams involving
two nucleons and one hyperon, as shown in Figs.~\ref{fig:LNN_pw},
\ref{fig:LNN_sw} and \ref{fig:LNN_d}. The first two diagrams correspond
to $P$-wave and $S$-wave 2$\pi$-exchange. The last diagram
represents a dispersive contribution associated with the  
medium modifications of the intermediate-state potentials for
the $\Sigma$, $N$, $\Delta$ due to the presence of 
the second nucleon. This term includes short-range contributions 
and it is expected to be repulsive due to the suppression mechanism 
associated with the $\Lambda N$-$\Sigma N$
coupling~\cite{Bodmer:1971,Rozynek:1979}.

As reported in Ref.~\cite{Usmani:2008},
the three-body potential $v_{\lambda ij}$ can be conveniently
decomposed in the 2$\pi$-exchange contributions
$v^{{2\pi}}_{\lambda ij}=v^{2\pi,P}_{\lambda
ij}+v^{2\pi,S}_{\lambda ij}$ (Figs.~\ref{fig:LNN_pw}
and \ref{fig:LNN_sw}) and the spin-dependent dispersive term $v_{\lambda
ij}^{D}$ (Fig.~\ref{fig:LNN_d}) as follows:
\begin{align}
	\hspace{-0.19cm}v_{\lambda ij}^{2\pi,P}&=-\frac{C_P}{6}
		\Bigl\{X_{i\lambda}\,,X_{\lambda j}\Bigr\}\,{\bm\tau}_{i}\cdot{\bm\tau}_{j}\;,\label{eq:V_LNN_P}\\[0.7em]
	\hspace{-0.19cm}v_{\lambda ij}^{2\pi,S}&=
		C_S\,Z\left(r_{\lambda i}\right)Z\left(r_{\lambda j}\right)\,
		{\bm\sigma}_{i}\cdot\hat{\bm r}_{i\lambda}\,
		{\bm\sigma}_{j}\cdot\hat{\bm r}_{j\lambda}\,{\bm\tau}_{i}\cdot{\bm\tau}_{j}\;,\label{eq:V_LNN_S}\\[0.7em]
	\hspace{-0.19cm}v_{\lambda ij}^{D}&=W_D\,
		T_{\pi}^{2}\left(r_{\lambda i}\right)T^{2}_{\pi}\left(r_{\lambda j}\right)
		\!\!\bigg[1+\frac{1}{6}{\bm\sigma}_\lambda\!\cdot\!\left({\bm\sigma}_{i}+{\bm\sigma}_{j}\right)\bigg]\;.\label{eq:V_LNN_D}
\end{align}
The function $T_\pi(r)$ is the same as in Eq.~(\ref{eq:T_pi}), 
while the $X_{\lambda i}$ and $Z(r)$ are defined by
\begin{align}
	X_{\lambda i}&=Y_{\pi}(r_{\lambda i})\;\bm\sigma_{\lambda}\cdot\bm\sigma_{i}+
	T_{\pi}(r_{\lambda i})\;S_{\lambda i}\;,\\[0.7em]
	Z(r)&=\frac{\mu_\pi r}{3} \Bigl[Y_\pi(r)-T_\pi(r)\Bigr]\;,
\end{align}
where
\begin{align}
	Y_\pi(r)=\frac{\e^{-\mu_\pi r}}{\mu_\pi r}\Bigl(1-\e^{-cr^2}\Bigr)
	\label{eq:Y_pi}
\end{align}
is the regularized Yukawa potential and $S_{\lambda i}$ is the 
same tensor operator as in Eq.~(\ref{eq:S_ij}).
The range of parameters $C_P$, $C_S$ and $W_D$ can be
found in Table~\ref{tab:parLN+LNN}. 
It is important to note that the three-body $\Lambda NN$ interaction has been
used in Variational Monte Carlo calculations for single $\Lambda$~hypernuclei 
($^3_\Lambda$H~\cite{Bodmer:1988,Shoeb:1999}, 
$^4_\Lambda$H and $^4_\Lambda$He~\cite{Bodmer:1985,Bodmer:1988,Shoeb:1999,Sinha:2002}, $^5_\Lambda$He~\cite{Bodmer:1988,Shoeb:1999,Usmani:1995_3B,Usmani:1999,Sinha:2002,Usmani:2003,Usmani:2006,Usmani:2008}, 
$^9_\Lambda$Be~\cite{Bodmer:1984,Shoeb:1998}, 
$^{13}_{~\Lambda}$C~\cite{Bodmer:1984}, 
$^{17}_{~\Lambda}$O~\cite{Usmani:1995,Usmani:1995_3B}) 
and double $\Lambda$~hypernuclei ($^{\;\;\,4}_{\Lambda\Lambda}$H, 
$^{\;\;\,5}_{\Lambda\Lambda}$H, $^{\;\;\,5}_{\Lambda\Lambda}$He~\cite{Shoeb:2004} and $^{\;\;\,6}_{\Lambda\Lambda}$He~\cite{Shoeb:2004,Usmani:2004,Usmani:2006_He6LL}), 
but no unique set of parameters has been set so far.

We stress the fact that, unlike to the nucleon sector, both
the two- and three-body hyperon-nucleon interactions are of the same
2$\pi$-exchange order. In addition, the mass of the intermediate excited
state $\Sigma$ compared to the $\Lambda$ is much smaller than in the
pure nucleonic case, where the difference between the nucleon and the
$\Delta$ resonance is much larger. The $\Lambda NN$ interaction should
therefore be considered necessary in addition to the $\Lambda N$ force
in any consistent theoretical calculation involving a $\Lambda$.
\begin{table}[ht]
\caption[]{Parameters of the $\Lambda N$ and $\Lambda NN$
interaction~(see~\cite{Usmani:2008} and references therein).  
For $C_P$ and $W_D$ the variational allowed range is shown. The 
value of the charge-symmetry breaking parameter $C_\tau$ is from
Ref.~\cite{Usmani:1999}.\label{tab:parLN+LNN}}
\begin{ruledtabular}
\begin{tabular}{ccc}
 Constant      & Value            & Unit      \\
\hline
		$W_c$		  & 2137             & MeV       \\
		$\bar r$	  & 0.5              & fm        \\
		$a$			  &	0.2              & fm   \\
		$v_s$		  &	6.33, 6.28       & MeV       \\
		$v_t$		  & 6.09, 6.04       & MeV       \\
		$\bar v$	  & 6.15(5)          & MeV       \\
		$v_\sigma$	  & 0.24             & MeV       \\
		$c$			  &	2.0              & fm$^{-2}$ \\
		$\varepsilon$ & $0.1\div0.38$    & --        \\
		$C_\tau$      & $-0.050(5)$      & MeV       \\
		$C_P$         &	$0.5\div2.5$     & MeV       \\
		$C_S$         &	$\sim 1.5$       & MeV       \\
		$W_D$         &	$0.002\div0.058$ & MeV       \\
\end{tabular}
\end{ruledtabular}
\end{table}

\subsection{Hyperon-Hyperon interaction}
\label{subsec:YY}

For the $\Lambda\Lambda$ potential, we follow the guide lines
adopted in the three- and four-body cluster models for double
$\Lambda$~hypernuclei~\cite{Hiyama:1997,Hiyama:2002}, which were
also used in Faddeev-Yakubovsky calculations for light double
$\Lambda$~hypernuclei~\cite{Filikhin:2002} and in variational
calculations on $^{\;\;\,4}_{\Lambda\Lambda}$H~\cite{Shoeb:2004,Shoeb:2005}, 
$^{\;\;\,5}_{\Lambda\Lambda}$H and 
$^{\;\;\,5}_{\Lambda\Lambda}$He~\cite{Shoeb:2004,Shoeb:2007} and 
$^{\;\;\,6}_{\Lambda\Lambda}$He~\cite{Usmani:2004,Usmani:2006_He6LL,Shoeb:2004,Shoeb:2007}, 
with different parametrizations.

The employed effective interaction is a low-energy phase equivalent
Nijmegen interaction represented by a sum of three Gaussians:
\begin{align}
	&v_{\lambda\mu}=\sum_{k=1}^{3}\left(v_0^{(k)}+v_\sigma^{(k)}\,
	{\bm\sigma}_\lambda\cdot{\bm\sigma}_\mu\right)
	\e^{-\mu^{(k)}r_{\lambda\mu}^2}\;. \label{eq:V_LL} 
\end{align}
The most recent parametrization of the potential (see
Table~\ref{tab:parLL}), has been fitted to
simulate the $\Lambda\Lambda$ sector of the Nijmegen F (NF)
interaction~\cite{Nagels:1979,Maessen:1989,Rijken:1999}. The NF 
model is the
simplest among the Nijmegen models with a scalar nonet, which seems
to be more appropriate than the versions including only a scalar
singlet in order to reproduce the weak binding energy indicated
by the NAGARA event~\cite{Takahashi:2001}. The components $k=1,2$
of the above Gaussian potential are determined so as to simulate the
$\Lambda\Lambda$ sector of NF and the strength of the part for $k=3$
is adjusted so as to reproduce the $^{\;\;\,6}_{\Lambda\Lambda}$He
NAGARA experimental double $\Lambda$~separation energy of $7.25\pm
0.19^{+0.18}_{-0.11}$~MeV.  In 2010, Nakazawa reported a new, more
precise determination of $B_{\Lambda\Lambda}=6.93\pm0.16$~MeV for
$^{\;\;\,6}_{\Lambda\Lambda}$He~\cite{Nakazawa:2010}, obtained via the
$\Xi^-$ hyperon capture at rest reaction in a hybrid emulsion.  This value
has been recently revised to $B_{\Lambda\Lambda}=6.91\pm0.16$~MeV by
the E373 (KEK-PS) Collaboration~\cite{Ahn:2013}.  No references were
found about the refitting of the $\Lambda\Lambda$ Gaussian potential on
the more recent experimental result, which is in any case compatible
with the NAGARA event. We therefore use the original parametrization
of Ref.~\cite{Hiyama:2002}.
\begin{table}[ht]
\caption[]{Parameters for the one-boson exchange potential simulating 
the $\Lambda\Lambda$ interaction. Depths $v_0^{(k)}$ and $v_\sigma^{(k)}$ 
[MeV] for each size parameter $\mu^{(k)}$ [fm$^{-2}$]~\cite{Hiyama:2002}.
\label{tab:parLL}}
\begin{ruledtabular}
\begin{tabular}{cccc}
$\mu^{(k)}$      & 0.555  & 1.656  & 8.163 \\
\hline
$v_0^{(k)}$      & -10.67 & -93.51 & 4884  \\
$v_\sigma^{(k)}$ & 0.0966 & 16.08  & 915.8 \\
\end{tabular}
\end{ruledtabular}
\end{table}

\section{Method}
\label{sec:method}

\subsection{Auxiliary field DMC method for nuclei}
\label{subsec:AFDMCnuc}

The auxiliary field diffusion Monte Carlo (AFDMC) method was introduced
by Schmidt and Fantoni~\cite{Schmidt:1999} as an extension of the
usual diffusion Monte Carlo (DMC) method to deal in an efficient way
with spin/isospin-dependent Hamiltonians. The standard DMC projects out the
ground state of the system by starting from a trial wave function not
orthogonal to the true ground state. By sampling configurations of the
system in coordinate-spin-isospin space, the trial wave function is
propagated in imaginary-time $\tau$. Expectation values are computed
averaging over the sampled configurations in the $\tau\rightarrow\infty$
limit, for which the evolved state approaches the ground state of the
Hamiltonian.

In nuclear Hamiltonians, the potential contains quadratic spin and
isospin operators, so the many-body wave function cannot
be written as a product of single-particle, spin-isospin states. The
number of components in the propagated wave function grows exponentially with $A$
and thus it quickly becomes computationally intractable. Standard
DMC calculations for nuclei are in fact limited
up to 12 nucleons~\cite{Pieper:2005,Lusk:2010,Lovato:2013} or 16
neutrons~\cite{Gandolfi:2011}.

The idea of the AFDMC method consists in reducing the quadratic
spin-isospin operators into linear terms in the propagator. The starting point
is to recast the Argonne V6- and V4-type potentials in spin-isospin independent
and spin-isospin dependent components, the latter of the form
\begin{align}
	V_{NN}	&=\frac{1}{2}\sum_{i\ne j}\sum_\gamma\tau_{i\gamma}\; 
	A_{ij}^{[\tau]}\;\tau_{j\gamma} \nonumber \\[0.2em]
	&\,+\frac{1}{2}\sum_{i\ne j}\sum_{\alpha\beta}\sigma_{i\alpha}\; 
	A_{i\alpha,j\beta}^{[\sigma]}\;\sigma_{j\beta} 
	\label{eq:V_NN_AFDMC} \\[0.2em]
	&\,+\frac{1}{2}\sum_{i\ne j}\sum_{\alpha\beta\gamma}\tau_{i\gamma}\,\sigma_{i\alpha}\; 
	A_{i\alpha,j\beta}^{[\sigma\tau]}\;\sigma_{j\beta}\,\tau_{j\gamma}\;. \nonumber
\end{align}
The matrices $A$ are real and symmetric with zero
diagonals and contain proper combinations of the components of AV6 and AV4
(latin indices are used for the
nucleons, greek ones refer to the cartesian components of the operators)
\begin{align}
	A_{ij}^{[\tau]}&=v_2\left(r_{ij}\right)\;,\nonumber\\[0.5em]
	A_{i\alpha,j\beta}^{[\sigma]}&=v_3\left(r_{ij}\right)\delta_{\alpha\beta}
	+v_5\left(r_{ij}\right)\left(3\,\hat r_{ij}^\alpha\,\hat r_{ij}^\beta-\delta_{\alpha\beta}\right)\;,\label{eq:A_NN}\\[0.5em]
	A_{i\alpha,j\beta}^{[\sigma\tau]}&=v_4\left(r_{ij}\right)\delta_{\alpha\beta}
	+v_6\left(r_{ij}\right)\left(3\,\hat r_{ij}^\alpha\,\hat r_{ij}^\beta-\delta_{\alpha\beta}\right)\;.\nonumber
\end{align}
By diagonalizing such matrices it is
possible to write the quadratic operators of Eq.~(\ref{eq:op}) in terms
of the eigenvectors of the matrices $A$. In the $\sigma$ channel we define
for example
\begin{align}
	\mathcal O_n^{[\sigma]}=\sum_{j\beta}\sigma_{j\beta}\,\psi_{n,j\beta}^{[\sigma]}\;,
\end{align}
where
\begin{align}
	\sum_{j\beta} A_{i\alpha,j\beta}^{[\sigma]}\,\psi_{n,j\beta}^{[\sigma]}=
	\lambda_n^{[\sigma]}\,\psi_{n,i\alpha}^{[\sigma]}\;.
\end{align}
Given the nucleon-nucleon spin-isospin dependent interaction in this
form, we can write the imaginary time propagator by means of the
Hubbard-Stratonovich (HS) transformation
\begin{equation}
	\e^{-\frac{1}{2}\tau\,\lambda_n\,\left(\mathcal O_n\right)^2}
	=\frac{1}{\sqrt{2\pi}}\int\!dx_n\,\e^{-\frac{x_n^2}{2}}
	\e^{\sqrt{-\tau\lambda_n}\,x_n\,\mathcal O_n}\;.
	\label{eq:HS}
\end{equation}
The newly introduced $x_n$ variables, called \emph{auxiliary fields},
are sampled to evaluate the integral of Eq.~(\ref{eq:HS}). The linearized
propagator has the effect of rotating the spin-isospin components of each
single nucleon. This eventually recovers the action of the quadratic
spin-isospin operators on the trial wave function containing all the
possible good spin-isospin states.  The procedure described reduces the
dependence to the number of operations needed to evaluate the trial wave
function from exponential to polynomial in the number of nucleons. The price
to pay is the additional computational cost due to the diagonalization
of the $A$ matrices and the sampling of the integral over auxiliary
fields. In any case, there is a net gain in computational time, 
the total number of AFDMC operations being at most proportional to~$A^3$.

The details of the AFDMC algorithm
for nuclei and neutron matter can be found in
Refs.~\cite{Gandolfi:2006,Gandolfi:2007,Gandolfi:2009},
where the adopted nuclear wave function, the computation of expectation
values, and the approximations used to overcome the Fermion sign
problem are discussed in detail.

\subsection{Auxiliary field DMC method for hypernuclei}
\label{subsec:AFDMChyp}

\emph{Hypernuclear wave function.}
\label{subsubsec:wave}
The starting point of all DMC methods is the set up of the trial wave
function.  The $\Lambda$~particle is distinguishable from the nucleons
so, to describe single and double $\Lambda$~hypernuclei, we write
a factorized trial wave function of the form
\begin{align}
	\psi_T(R,S)=\psi_T^N(R_N,S_N)\,\psi_T^{\Lambda}(R_\Lambda,S_\Lambda)\;,
	\label{eq:psiT}
\end{align}
where $R=\{R_N,R_\Lambda\}$ and $S=\{S_N,S_\Lambda\}$
with $R_p=\{\bm r_1,\ldots,\bm r_{\mathcal N_p}\}$ and
$S_p=\{ s_1,\ldots, s_{\mathcal N_p}\}$, $p=N,\Lambda$.
The two components of $\psi_T$ are chosen of the same form of
Refs.~\cite{Gandolfi:2007}
\begin{align}
	\!\!\!\psi_T^p\left(R_p,S_p\right)=\!\Bigg[\prod_{i<j}f_{ij}\Bigg]_p\!
	\mathcal A\Bigg[\prod_i\varphi_i\left(\bm r_i-\bm r_{CM}, s_i\right)\!\Bigg]_p,\!
	\label{eq:psi_T_p}
\end{align}
where $\mathcal A$ is an antisymmetrization operator, $\bm r_{CM}$ is the
center of mass of the hypernucleus, and $\varphi_i$ are single-particle space
and spin-isospin orbitals, built from combinations of radial functions,
spherical harmonics and spinors.  Radial orbitals are the solutions of
the self-consistent Hartree-Fock problem with the Skyrme effective
interactions of Ref.~\cite{Bai:1997}. For the $\Lambda$~particle, we
assume the neutron $1s_{1/2}$ radial function.  Spinors are defined
as four-component complex vectors for the nucleons and two components
complex vectors for the $\Lambda$~particles
\begin{align}
	\!\!\! s_i^{\,N}\!&=\!\left(
	\begin{array}{c}
		a_i\\
		b_i\\
		c_i\\
		d_i
	\end{array}
	\right)\!=a_i|p\uparrow\rangle+b_i|p\downarrow\rangle+
	c_i|n\uparrow\rangle+d_i|n\downarrow\rangle\;,\label{eq:spinor}\\[0.5em]
	\!\!\! s_i^{\,\Lambda}\!&=\!\left(
	\begin{array}{c}
		u_i\\
		v_i
	\end{array}
	\right)\!=u_i|\Lambda\uparrow\rangle+v_i|\Lambda\downarrow\rangle\;.
\end{align}
The functions $f_{ij}$ are symmetric and spin independent Jastrow correlation
functions, solutions of the Schr\"odinger-like equation for $f_{ij}(r<d)$
\begin{align}
	-\frac{\hbar^2}{2\mu_{ij}}\nabla^2 f_{ij}(r)+\eta\,v_{ij}^c(r)f_{ij}(r)=\xi f_{ij}(r)\;,
\end{align}
where $v_{ij}^c(r)$ is the spin independent part of the two-body interaction, 
$\mu_{ij}=m_p/2$ the reduced mass of the pair, and $\eta$ and the healing 
distance $d$ are variational parameters. For distances $r\ge d$, we 
impose $f_{ij}(r)=1$.
The role of $f_{ij}$ functions is to include the short-range 
correlations in the trial wave function. In the AFDMC algorithm the
effect is simply a reduction of the overlap between pairs of particles,
with the reduction of the energy variance. Since there is no change
in the phase of the wave function, the $f_{ij}$ do not influence the computed
energy value in projection methods.

With the presented wave function, we consider nucleons and hyperons
as distinct particles.  In this way, it is not possible to include
the $\Lambda N$ exchange term of Eq.~(\ref{eq:V_LN}) directly in the
propagator, because it mixes hyperon and nucleon states. The complete
treatment of this factor would require an enlarged hyperon-nucleon isospin
space, which at present has not yet been  developed. A perturbative
analysis of the $v_0(r)\varepsilon(\mathcal P_x-1)$ term is however
possible as described in Ref.~\cite{Lonardoni:2013_PRC(R)}.

\emph{Algorithm.}
\label{subsubsec:algorithm}
The idea of the standard AFDMC method can be easily extended to
$\Lambda$~hypernuclear systems with the interactions described in
Sects.~\ref{subsec:YN} and \ref{subsec:YY}. 
Consider the hypernuclear potentials of Eqs.~(\ref{eq:V_LN}),
(\ref{eq:V_CSB}), (\ref{eq:V_LNN_P})-(\ref{eq:V_LNN_D}) and (\ref{eq:V_LL}), 
and assume the notations
\begin{align}
	T_{\lambda i}&=T_{\pi}(r_{\lambda i})\;,\nonumber \\[0.5em]
	Y_{\lambda i}&=Y_{\pi}(r_{\lambda i})\;, \\[0.5em]
	Q_{\lambda i}&=Y_{\lambda i}-T_{\lambda i}\;.\nonumber
\end{align}
In analogy with the nucleon-nucleon $A$ matrices of Eqs.~(\ref{eq:A_NN}), 
we can define the following hyperon-nucleon and hyperon-hyperon
matrices (Greek $\lambda$, $\mu$ indices indicates the $\Lambda$~particles)
\begin{align}
	B_{\lambda i}^{[\sigma]}&=\frac{1}{4}v_\sigma T_{\lambda i}\;,\\[0.5em]
	C_{\lambda i}^{[\sigma]}&=\frac{1}{3}W_D\!
	\sum_{j,j\ne i}T^2_{\lambda i}\,T^2_{\lambda j}\;,\\[0.5em]
	C_{i\alpha,j\beta}^{[\sigma\tau]}&=\sum_\lambda\Bigg\{\!-\frac{1}{3} 
	C_P Q_{\lambda i}Q_{\lambda j}\delta_{\alpha\beta}
	-C_P Q_{\lambda j}T_{\lambda i}\,\hat r_{i\lambda}^{\,\alpha}\,
	\hat r_{i\lambda}^{\,\beta} \nonumber \\[0.2em]
	&\quad-C_P Q_{\lambda i}T_{\lambda j}\,
	\hat r_{j\lambda}^{\,\alpha}\,\hat r_{j\lambda}^{\,\beta}
	+\!\left[\frac{1}{9} C_S \mu_\pi^2 Q_{\lambda i} 
	Q_{\lambda j}\,|r_{i\lambda}||r_{j\lambda}|\right. \nonumber \\[0.2em]
	&\quad\left.-3\,C_PT_{\lambda i}T_{\lambda j } 
	\left({\sum_\delta}\,\hat r_{i\lambda}^{\,\delta}\,
	\hat r_{j\lambda}^{\,\delta}\right)\right]\,\hat r_{i\lambda}^{\,\alpha}\,
	\hat r_{j\lambda}^{\,\beta}\Bigg\}\;,\\[0.5em]
	D_{\lambda\mu}^{[\sigma]}&=
	\sum_{k=1}^{3}v_\sigma^{(k)}\e^{-\mu^{(k)}r_{\lambda\mu}^2}\;.
\end{align}
In such a way it is possible to
recast the $\Lambda N$, $\Lambda NN$ and $\Lambda\Lambda$ interactions
so that they contain at most two-body operators in the hyperon-nucleon
extended space
\begin{align}
	V_{\Lambda N}&=\sum_{\lambda i}\sum_\alpha\sigma_{\lambda\alpha}
	\; B_{\lambda i}^{[\sigma]}\;\sigma_{i\alpha}+\widetilde V_{\Lambda N}\;,
	\label{eq:V_LN_2_AFDMC} \\[0.5em]
	V^{2\pi}_{\Lambda NN}&=\frac{1}{2}\sum_{i\ne j}
	\sum_{\alpha\beta\gamma}\tau_{i\gamma}\,\sigma_{i\alpha}
	\; C_{i\alpha,j\beta}^{[\sigma\tau]}\;\sigma_{j\beta}\,\tau_{j\gamma}\;,
	\label{eq:V_LNN_2pi_AFDMC} \\[0.5em]
	V^{D}_{\Lambda NN}&=
	\frac{1}{2}\sum_{\lambda i}\sum_\alpha\sigma_{\lambda\alpha}\; 
	C_{\lambda i}^{[\sigma]}\;
	\sigma_{i\alpha}+\widetilde V_{\Lambda NN}\;,
	\label{eq:V_LNN_D_AFDMC} \\[0.5em]
	V_{\Lambda\Lambda}&=\frac{1}{2}\sum_{\lambda\ne\mu}
	\sum_{\alpha}\sigma_{\lambda\alpha}\; D_{\lambda\mu}^{[\sigma]}\;
	\sigma_{\mu\alpha}+\widetilde V_{\Lambda\Lambda}\;,
	\label{eq:V_LL_AFDMC}	
\end{align}
where $\widetilde V_{\Lambda N}$, 
$\widetilde V_{\Lambda NN}$ and 
$\widetilde V_{\Lambda\Lambda}$ include all
the spin-isospin independent and the linear $\mathcal P_x$, $\tau_i^z$ terms
\begin{align}
	\widetilde V_{\Lambda N}&=\sum_{\lambda i}v_{0}(r_{\lambda i})(1-\varepsilon)
		+\sum_{\lambda i}v_{0}(r_{\lambda i})\,\varepsilon\,\mathcal P_x \nonumber\\[0.2em]
		&\,+\sum_{\lambda i} C_\tau\,T_\pi^2(r_{\lambda i})\,\tau_i^z\;,\nonumber\\[0.5em]
	\widetilde V_{\Lambda NN}&=\sum_{\lambda,i<j}W_D\,
		T_{\pi}^{2}\left(r_{\lambda i}\right)T^{2}_{\pi}\left(r_{\lambda j}\right)\;,\\[0.5em]
	\widetilde V_{\Lambda\Lambda}&=\sum_{\lambda<\mu}\sum_{k=1}^{3}v_0^{(k)}
		\e^{-\mu^{(k)}r_{\lambda\mu}^2}\;.\nonumber
\end{align}
All the remaining terms of 
Eqs.~(\ref{eq:V_LN_2_AFDMC})-(\ref{eq:V_LL_AFDMC})
consist of two-body spin-isospin operators  of exactly the same type as
those of Eqs.~(\ref{eq:V_NN_AFDMC}). 

The algorithm follows then the nuclear version with the sampling
of the coordinates, which now contains also the $\Lambda$ particles,
and of the auxiliary fields, one for each linearized operator.
The application of the propagator of Eq.~(\ref{eq:HS}) has
the effect of rotating the spinors of nucleons and $\Lambda$s.
The strategy adopted in order to control the Fermion sign
problem and reduce the variance of the estimators is the same of
Refs.~\cite{Gandolfi:2006,Gandolfi:2007,Gandolfi:2009},
with a straightforward extension to the enlarged hyperon-nucleon space.

The structure of the AFDMC algorithm for $\Lambda$~hypernuclei closely
follows the usual AFDMC procedure
\begin{enumerate}
\item Sample the nucleons and $\Lambda$'s positions, spins and isospins from
$|\psi_T(R,S)|^2$, using the Metropolis Monte Carlo method.

\item\label{item:1} Propagate the spatial degrees of freedom as in the
usual diffusion Monte Carlo with a drifted Gaussian for a small time step.

\item For each set of generalized coordinates (\emph{walker}), build
and diagonalize the potential matrices $A$, $B$, $C$ and $D$.

\item Loop over the eigenvectors, sampling the corresponding auxiliary
fields and rotating the spinors.

\item\label{item:2} Apply the fixed phase prescription and evaluate
the estimator contributions to averages for the calculation of expectation
values.

\item Iterate from \ref{item:1} to \ref{item:2} as long as necessary
until convergence of the energy is reached.  Error bars on expectation
values are then estimated by means of block averages and the analysis
of auto-correlations on data blocks.  
\end{enumerate}

\section{Results and discussion}
\label{sec:results}

\subsection{Single $\Lambda$~hypernuclei}
\label{subsec:singleL}

A direct comparison of energy calculations with experimental results is
given for the $\Lambda$~separation energy, defined as
\begin{align}
	&B_{\Lambda}\left(\,^A_\Lambda\text{Z}\,\right)=
	E\left(\,^{A-1}\text{Z}\,\right)-E\left(\,^A_\Lambda\text{Z}\,\right)\;,
	\label{eq:BL}
\end{align}
where $E$ is the energy of the system, i.e. the ground state expectation
value of the Hamiltonian,
\begin{align}
	E(\kappa)=\frac{\langle\psi_\kappa^0|H_\kappa|\psi_\kappa^0\rangle}
	{\langle\psi_\kappa^0|\psi_\kappa^0\rangle}\;,\quad\quad\kappa={\rm nuc},{\rm hyp}\;.
\end{align}
The computation of $B_{\Lambda}$ thus involves the calculation of the 
energy of the nucleus $^{A-1}\text{Z}$ and the corresponding
hypernucleus $^A_\Lambda\text{Z}$. The nuclear wave function is the same
as in Eq.~(\ref{eq:psi_T_p}) with the spinor of Eq.~(\ref{eq:spinor}).
As reported in Refs.~\cite{Lonardoni:2013_HYP2012,Lonardoni:2013_PRC(R)}, the
$\Lambda$~separation energy is not sensitive to the details of the nuclear
interactions. On the grounds of this observation, we adopt the nuclear
potential AV4' for both nuclei and hypernuclei in the present work.
This choice makes AFDMC calculations less expensive and more stable. The
resulting absolute binding energies are not comparable with experimental
results, but the estimated $B_\Lambda$ is in any case realistic.

In our previous work~\cite{Lonardoni:2013_PRC(R)}, we tackled the problem
of hyperon-nucleon interaction by studying closed-shell single
$\Lambda$~hypernuclei with the inclusion of two- and three-body
$\Lambda$-nucleon forces. The set of parameters for the $\Lambda NN$
potential was originally taken from Ref.~\cite{Usmani:1995_3B}, being
the choice that made the variational $B_\Lambda$ for $_\Lambda^5$He and
$^{17}_{~\Lambda}$O compatible with the expected results. It reads
\begin{equation*}
	\hypertarget{par_I}{(\text{I})}\phantom{I}\quad
	\left\{
	\begin{array}{rcll}
		C_P&\!=\!&0.60& \!\text{MeV}\\
		C_S&\!=\!&0.00& \!\text{MeV}\\
		W_D&\!=\!&0.015&\!\text{MeV}
	\end{array}
	\right.
\end{equation*}
The main outcome of the study is that the saturation property of the
$\Lambda$ binding energy is reproduced only with the inclusion of the
$\Lambda NN$ interaction.  However, with the given parametrization, only
a qualitative accord with the experimental results is obtained. Thus,
a refitting procedure for the three-body hyperon-nucleon interaction is
needed.

As reported in Ref.~\cite{Usmani:2008}, the $C_S$
parameter can be estimated by comparing the $S$-wave term of
Eq.~(\ref{eq:V_LNN_S}) with the Tucson-Melbourne model of the
$NNN$ force reported in Ref.~\cite{Pieper:2001}. We take the same
$C_S=1.50$~MeV value, in order to reduce the
number of fitting parameters.  This
choice is justified because the $S$-wave component of the three-body $\Lambda NN$
interaction is sub-leading.
We indeed verified that a change in the $C_S$ value yields a variation of the total 
energy within statistical error bars and definitely much smaller than the variation 
in energy due to a change of the $C_P$ and $W_D$ parameters.
\begin{figure}[ht]
	\centering
	\includegraphics[width=\linewidth]{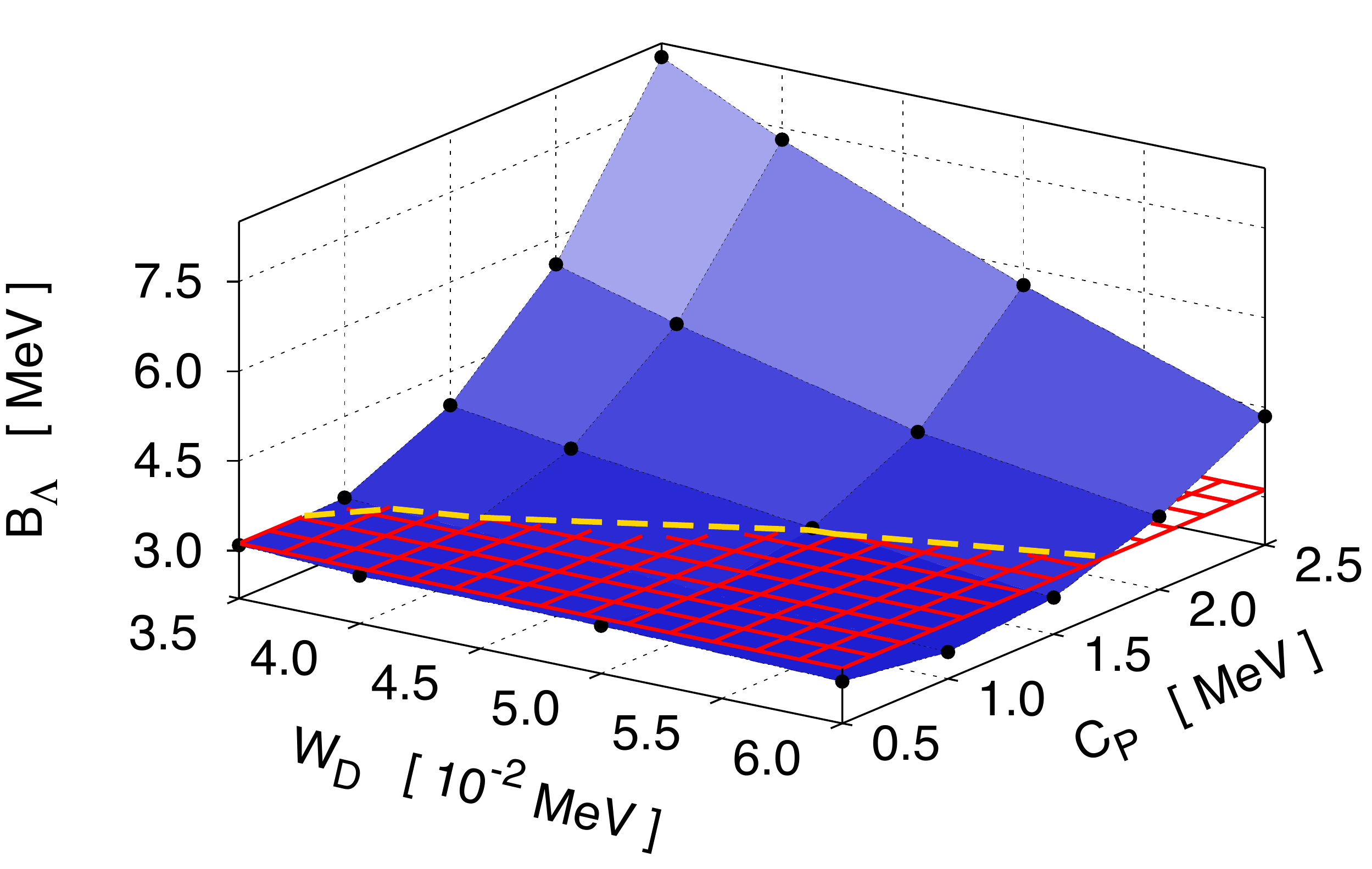}
	\caption[]{(Color online) $\Lambda$~separation energy for $^5_\Lambda$He as a function
		of strengths $W_D$ and $C_P$
		of the three-body $\Lambda NN$ interaction.  The red grid represents the
		experimental $B_\Lambda=3.12(2)$~MeV~\cite{Juric:1973}. The dashed yellow
		curve is the interception between the expected result and the $B_\Lambda$
		surface in the $W_D-C_P$
		parameter space.  Statistical error bars on AFDMC results (solid black 
		dots) are of the order of $0.10\div0.15$~MeV.}
	\label{fig:Wd-Cp_3D}
\end{figure}
\begin{figure}[ht]
	\centering
	\includegraphics[width=\linewidth]{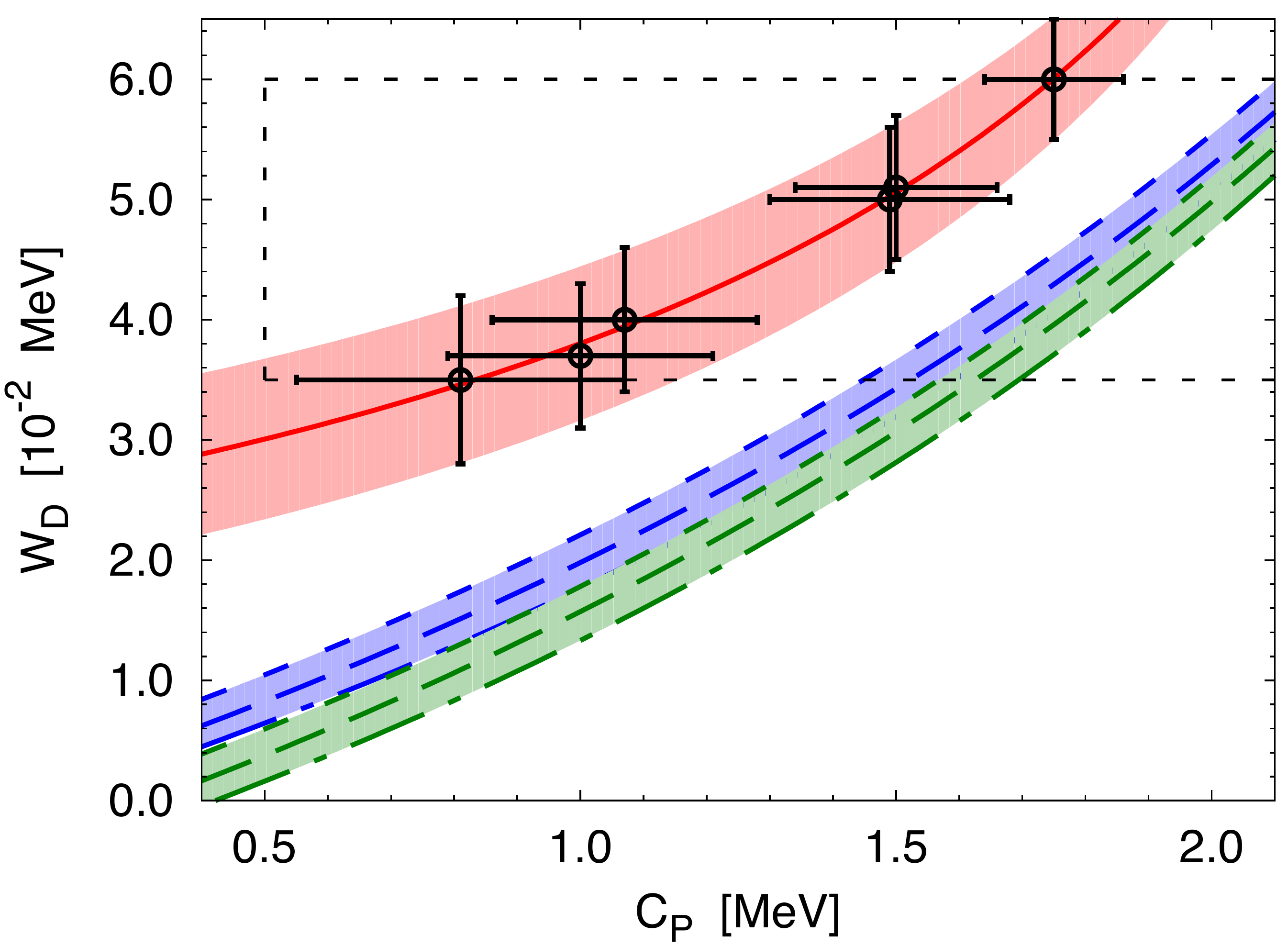}
	\caption[]{(Color online) Projection of Fig.~\ref{fig:Wd-Cp_3D} on the
		$W_D-C_P$ plane. Error
		bars come from a realistic conservative estimate of the uncertainty
		in the determination of the parameters due to the statistical errors
		of the Monte Carlo calculations. Blue and green dashed, long-dashed
		and dot-dashed lines (lower curves) are the variational results of
		Ref.~\cite{Usmani:2008} for different $\varepsilon$ and $\bar v$
		(two-body $\Lambda N$ potential). The dashed box corresponds to the
		parameter domain of Fig.~\ref{fig:Wd-Cp_3D}. Black dots and the red
		band (upper curve) are the projected interception describing
		the possible set of parameters reproducing the experimental~$B_\Lambda$.}
	\label{fig:Wd-Cp_2D}
\end{figure}

In Fig.~\ref{fig:Wd-Cp_3D} we report the systematic study of the
$\Lambda$~separation energy of $_\Lambda^5$He as a function of
both $W_D$ and $C_P$.
Solid black dots are the AFDMC results. The red grid represents the
experimental $B_\Lambda=3.12(2)$~MeV~\cite{Juric:1973}.  The dashed
yellow curve follows the set of parameters reproducing the
expected $\Lambda$~separation energy.  The same curve is also reported
in Fig.~\ref{fig:Wd-Cp_2D} (red band with black dots and
error bars), that is a projection of Fig.~\ref{fig:Wd-Cp_3D} on
the $W_D-C_P$ plane.
The dashed box represents the $W_D$ and
$C_P$ domain of the previous picture.  For
comparison, the variational results of Ref.~\cite{Usmani:2008}
are also reported. Green curves are the results for $\bar v=6.15$~MeV
and $v_\sigma=0.24$~MeV, blue ones for $\bar v=6.10$~MeV and
$v_\sigma=0.24$~MeV. Dashed, long-dashed and dot-dashed lines correspond
respectively to $\varepsilon=0.1$, $0.2$ and $0.3$.

In our calculations, we have not considered different combinations for
the parameters of the two-body $\Lambda N$ interaction, focusing on
the three-body part. We have thus kept fixed $\bar v$ and $v_\sigma$
to the same values of the green curves of Fig.~\ref{fig:Wd-Cp_2D}
(see Table~\ref{tab:parLN+LNN} for the detailed list of constants).
Moreover, in the present work we have set $\varepsilon=0$ for all the
studied hypernuclei due to the impossibility of exactly including the 
$\mathcal P_x$ exchange operator in the propagator. However, from a
perturbative analysis, the net effect of the $v_0(r)\varepsilon(\mathcal
P_x -1)$ term on the hyperon separation energy within the statistical
errors of the Monte Carlo calculation, seems to be the same as a slight change in
the strength of the central $\Lambda N$ potential.

\begin{figure}[hb]
	\centering
	\includegraphics[width=\linewidth]{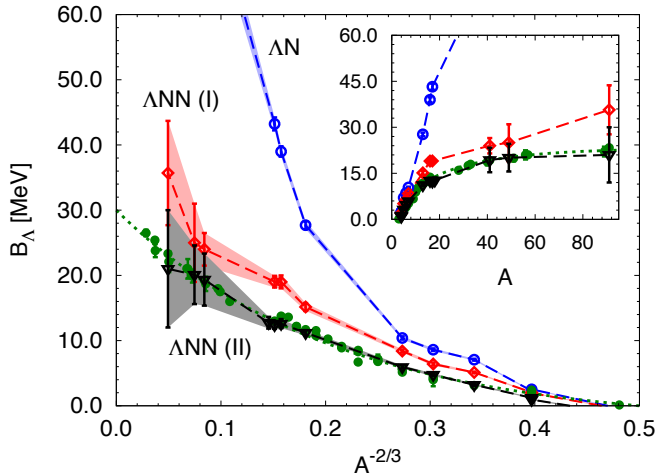}
	\caption[]{(Color online) $\Lambda$~separation energy as a function of $A^{-2/3}$.
		Solid green dots~(dashed curve) are the available $B_\Lambda$
		experimental or semiempirical values.  Empty blue dots~(upper
		banded curve) refer to the AFDMC results for the two-body $\Lambda N$
		interaction alone.  Empty red diamonds~(middle banded curve) and empty
		black triangles~(lower banded curve) are the results with the inclusion
		also of the three-body hyperon-nucleon force, respectively for the set
		of parameters (\hyperlink{par_I}{I}) and (\hyperlink{par_II}{II}). In
		the inset, the same data plotted as a function of $A$.}
	\label{fig:BL-A23}
\end{figure}

Starting from the analysis of the results in the
$W_D-C_P$ space for
$_\Lambda^5$He, we performed simulations for the next closed-shell
hypernucleus $^{17}_{~\Lambda}$O.  Using the parameters in the red band
of Fig.~\ref{fig:Wd-Cp_2D} we identified a parametrization able
to reproduce the experimental $B_\Lambda$ for both $_\Lambda^5$He and
$^{17}_{~\Lambda}$O at the same time, namely
\begin{equation*}
	\hypertarget{par_II}{(\text{II})}\quad
	\left\{
	\begin{array}{rcll}
		C_P&\!=\!&1.00& \!\text{MeV}\\
		C_S&\!=\!&1.50& \!\text{MeV}\\
		W_D&\!=\!&0.035&\!\text{MeV}\ .
	\end{array}
	\right.
\end{equation*}
Given the set (\hyperlink{par_II}{II}), the $\Lambda$~separation energy
of closed-shell and open-shell single $\Lambda$~hypernuclei has been
calculated in a mass range $3 \leq A \leq 91$.  The closed-shell
hypernuclei are the same of Ref.~\cite{Lonardoni:2013_PRC(R)}.  The results
are summarized in Fig.~\ref{fig:BL-A23}, where we report $B_\Lambda$
as a function of $A^{-2/3}$, and as a function of $A$ in the inset.
Solid green dots are the available experimental data, empty symbols the
AFDMC results.  
The blue curve is obtained using only the two-body hyperon-nucleon
interaction in addition to the nuclear AV4' potential.  The red
curve refers to the results for the same systems when also the
three-body $\Lambda NN$ interaction with the old set of parameters
(\hyperlink{par_I}{I}) is included.  The black lower curve shows the results obtained by including the three-body hyperon-nucleon interaction 
described by the new parametrization~(\hyperlink{par_II}{II}).
A detailed comparison between numerical results and experiments for the hyperon-separation energy
can be found in Table~\ref{tab:BL}.

\begin{table}[ht]
\caption[]{$\Lambda$~separation energies (in MeV) obtained using the
two-body plus three-body hyperon-nucleon interaction with the set of
parameters (\hyperlink{par_II}{II}). 
The results already include the CSB contribution.
In the last column, the expected $B_\Lambda$ values.
No experimental data for $A=17,18,41,49,91$ exists.
For $^{17}_{~\Lambda}$O the reference separation energy is a semiempirical value. 
For $A=41,49,91$ the experimental hyperon~binding energies are those of the nearest hypernuclei
$^{40}_{~\Lambda}$Ca, $^{51}_{~\Lambda}$V and $^{89}_{~\Lambda}$Y respectively.
\label{tab:BL}}
\begin{ruledtabular}
\begin{tabular}{ccc}
		System         & {AFDMC $B_\Lambda$} & {Expt. $B_\Lambda$} \\
\hline
		\hspace{-0.3em}$^3_\Lambda$H        &  -1.22(15)  &  0.13(5)  \hspace{1.2em}\cite{Juric:1973}    \\
		\hspace{-0.3em}$^4_\Lambda$H        &   0.95(9)   &  2.04(4)  \hspace{1.2em}\cite{Juric:1973}    \\
		\hspace{0.1em}$^4_\Lambda$He       &   1.22(9)   &  2.39(3)  \hspace{1.2em}\cite{Juric:1973}    \\
		\hspace{0.1em}$^5_\Lambda$He       &   3.22(14)  &  3.12(2)  \hspace{1.2em}\cite{Juric:1973}    \\
		\hspace{0.1em}$^6_\Lambda$He       &   4.76(20)  &  4.25(10) \hspace{0.7em}\cite{Juric:1973}    \\
		\hspace{0.1em}$^7_\Lambda$He       &   5.95(25)  &  5.68(28) \hspace{0.7em}\cite{Nakamura:2013} \\
		\hspace{-0.6em}$^{13}_{~\Lambda}$C  &  11.2(4)    & 11.69(12) \hspace{0.2em}\cite{Cantwell:1974} \\
		\hspace{-0.5em}$^{16}_{~\Lambda}$O  &  12.6(7)    & 12.42(41) \hspace{0.2em}\cite{Hashimoto:2006} \\
		\hspace{-0.5em}$^{17}_{~\Lambda}$O  &  12.4(6)    & 13.0(4)   \hspace{1.2em}\cite{Usmani:1995}   \\
		\hspace{-0.5em}$^{18}_{~\Lambda}$O  &  12.7(9)    & \hspace{-3.0em}---                           \\
		$^{41}_{~\Lambda}$Ca &  19(4)      & 18.7(1.1) \hspace{0.43em}\cite{Pile:1991}   \\
		$^{49}_{~\Lambda}$Ca &  20(5)      & 19.97(13) \hspace{0.18em}\cite{Hotchi:2001}   \\
		$^{91}_{~\Lambda}$Zr &  21(9)      & 23.11(10) \hspace{0.18em}\cite{Hotchi:2001}   \\
\end{tabular}
\end{ruledtabular}
\end{table}

For systems with $A\ge 5$, all the $\Lambda$~separation energies are
compatible with the expected results, where available.  For $A<5$
our results are more than 1~MeV off from experimental data. For
$^3_\Lambda$H, the $\Lambda$~separation energy is even negative, meaning
that the hypernucleus is less bound than the corresponding nucleus $^2$H.
We can ascribe this discrepancy to the lack of accuracy of our nucleonic 
wavefunction. Moreover, the
single particle orbitals might need to be changed when the $\Lambda$ is
added to the nucleus. This effect is expected to be much less important
for heavier hypernuclei where bulk effects dominate over the surface.  A study
of these systems within a few-body method might solve this issue.

The effect of the CSB potential has been studied for the $A=4$ mirror hypernuclei.  
As reported in Table~\ref{tab:CSB_A4}, without the CSB term there is no difference 
in the $\Lambda$ binding energy of $^4_\Lambda$H and $^4_\Lambda$He. When CSB 
is active, a splitting appears due to the different behavior of the $\Lambda
p$ and $\Lambda n$ channels. The strength of the difference $\Delta
B_\Lambda^{CSB}$ is independent on the parameters
of the three-body $\Lambda NN$ interaction and it is compatible with
the experimental result~\cite{Juric:1973}.

\begin{table}[htb]
\caption[]{$\Lambda$~separation energies (in MeV) for the $A=4$ mirror
$\Lambda$~hypernuclei with (fourth column) and without (third column)
the inclusion of the charge symmetry breaking term.
In the last column the difference in the separation energy induced by
the CSB interaction. First and second rows refer to different set of
parameters for the $\Lambda NN$ interaction, while the last row is the
experimental result.\label{tab:CSB_A4}}
\begin{ruledtabular}
\begin{tabular}{ccccc}
Parameters & System &   $B_\Lambda^{sym}$ & $B_\Lambda^{CSB}$
& {$\Delta B_\Lambda^{CSB}$} \\
\hline
		\multirow{2}{*}{set (\hyperlink{par_I}{I})}   
		& $^4_\Lambda$H\hspace{0.4em} & 1.97(11) &  1.89(9) & \multirow{2}{*}{0.24(12)} \\
		& $^4_\Lambda$He              & 2.02(10) &  2.13(8) &                           \\[0.8em]
		\multirow{2}{*}{set (\hyperlink{par_II}{II})} 
		& $^4_\Lambda$H\hspace{0.4em} & 1.07(8)  &  0.95(9) & \multirow{2}{*}{0.27(13)} \\
		& $^4_\Lambda$He              & 1.07(9)  &  1.22(9) &                           \\
		\midrule
		\multirow{2}{*}{Expt.~\cite{Juric:1973}} & $^4_\Lambda$H\hspace{0.4em} &  {---}   &  2.04(4) & \multirow{2}{*}{0.35(5)\phantom{0}}  \\
		& $^4_\Lambda$He              &  {---}   &  2.39(3) &                           \\
\end{tabular}
\end{ruledtabular}
\end{table}

\begin{table}[ht]
\caption[]{Difference (in MeV) in the hyperon separation energies induced by 
the CSB term for different hypernuclei. The fourth column reports the 
difference between the number of neutrons and protons. 
Results are obtained with the full two- plus three-body 
(set (\hyperlink{par_II}{II})) hyperon-nucleon interaction.
In order to reduce the errors, $\Delta B_\Lambda$ has been calculated 
by taking the difference between total hypernuclear binding energies, 
instead of the hyperon separation energies.\label{tab:CSB}}
\begin{ruledtabular}
\begin{tabular}{ccccc}
System  & $p$ & $n$ &   $\Delta_{np}$    & $\Delta B_\Lambda$ \\
\hline
		\hspace{-0.5em}$^4_\Lambda$H       & 1 & 2 &       $+1$         & $-0.12(8)$ \\[0.5em]     
		$^4_\Lambda$He                     & 2 & 1 &       $-1$         & $+0.15(9)$ \\       
		$^5_\Lambda$He                     & 2 & 2 & \hspace{0.78em}$0$ & $+0.02(9)$ \\       
		$^6_\Lambda$He                     & 2 & 3 &       $+1$         & $-0.06(8)$ \\       
		$^7_\Lambda$He                     & 2 & 4 &       $+2$         & $-0.18(8)$ \\[0.5em]
		\hspace{-0.7em}$^{16}_{~\Lambda}$O & 8 & 7 &       $-1$         & $+0.27(35)$ \\      
		\hspace{-0.7em}$^{17}_{~\Lambda}$O & 8 & 8 & \hspace{0.78em}$0$ & $+0.15(35)$ \\      
		\hspace{-0.7em}$^{18}_{~\Lambda}$O & 8 & 9 &       $+1$         & $-0.74(49)$ \\      
\end{tabular}
\end{ruledtabular}
\end{table}

The same CSB potential of Eq.~(\ref{eq:V_CSB}) has been included
in the study of hypernuclei for $A>4$.  In Table~\ref{tab:CSB},
the difference in the hyperon separation energies $\Delta
B_\Lambda=B_\Lambda^{CSB}-B_\Lambda^{sym}$
is reported for different hypernuclei up to $A=18$. The fourth
column shows the difference between the number of neutrons and protons
$\Delta_{np}=\mathcal N_n-\mathcal N_p$.  For the symmetric hypernuclei
$^5_\Lambda$He and $^{17}_{~\Lambda}$O the CSB interaction has no
effect, this difference being zero.  In the systems with neutron excess
($\Delta_{np}>0$), the effect of the CSB consists in decreasing the
hyperon separation energy compared to the charge symmetric case. When
$\Delta_{np}$ becomes negative, $\Delta B_\Lambda>0$ due to the attraction
induced by the CSB potential in the $\Lambda p$ channel, producing
more bound hypernuclei. These effects are in any case rather small
and they become almost negligible compared to the statistical errors on
$B_\Lambda$ when the number of baryons becomes large enough ($A>16$).

Single-particle densities can be computed in Monte Carlo calculations by 
considering the expectation value of the density operator
\begin{align}
	\hat\rho_\kappa(r)=\sum_{i}\delta(r-r_i)\quad\quad \kappa=N,\Lambda\;,
\end{align}
where $i$ is the single particle index running over nucleons for $\rho_N=\langle\hat\rho_N\rangle$ or hyperons for $\rho_\Lambda=\langle\hat\rho_\Lambda\rangle$. 
The correct estimator for positive defined operators $\mathcal O$ different from the total Hamiltonian is obtained starting from
the mixed DMC result and the variational one via the relation~\cite{Pieper:2008}
\begin{align}
	\!\!\langle\mathcal O\rangle_{real}=
	\frac{\langle\psi_0|\mathcal O|\psi_0\rangle}{\langle\psi_0|\psi_0\rangle}
	=\frac{\left(\frac{\langle\psi_T|\mathcal O|\psi_0\rangle}{\langle\psi_T|\psi_0\rangle}\right)^2}
	{\frac{\langle\psi_T|\mathcal O|\psi_T\rangle}{\langle\psi_T|\psi_T\rangle}}
	=\frac{\langle\mathcal O\rangle_{\text{DMC}}^2}{\langle\mathcal O\rangle_{\text{VMC}}}\;,
\end{align}
where $\psi_T$ is the trial wave function and $\psi_0$ the projected ground state 
wave function. Although easy to implement, the calculation of single-particle 
densities in the present version of the AFDMC algorithm suffers of two main issues.
On one hand, the employed trial wave function is too poor for variational 
calculations. The estimate of $\langle\mathcal O\rangle_{\text{VMC}}$ could 
be not accurate enough, introducing severe biases in the calculation of 
$\langle\mathcal O\rangle_{real}$. On the other hand, the employed $NN$ potential 
is too simplified to correctly describe the physics of nucleons in nuclei and 
hypernuclei, particularly for heavy systems. This lack of accuracy does not 
affect the calculation of the hyperon separation energy but could be important 
in the estimate of single particle densities. For these reasons we do not report 
here the results for nucleon and hyperon single-particle densities which will be 
presented in a future work in connection with a better trial wave function and 
a more realistic nucleon-nucleon interaction.

\subsection{Double $\Lambda$~hypernuclei}
\label{subsec:doubleL}

In the case of double $\Lambda$~hypernuclei, the interesting observables
we can have access with the AFDMC are the double $\Lambda$~separation
energy
\begin{align}
	&B_{\Lambda\Lambda}\left(\,^{~\,A}_{\Lambda\Lambda}\text{Z}\,\right)=
	E\left(\,^{A-2}\text{Z}\,\right)-E\left(\,^{~\,A}_{\Lambda\Lambda}\text{Z}\,\right)\;,
	\label{eq:BLL}
\end{align}
and the incremental $\Lambda\Lambda$ energy
\begin{align}
	&\Delta B_{\Lambda\Lambda}\left(\,^{~\,A}_{\Lambda\Lambda}\text{Z}\,\right)=
	B_{\Lambda\Lambda}\left(\,^{~\,A}_{\Lambda\Lambda}\text{Z}\,\right)
	-2 B_\Lambda\left(\,^{A-1}_{\quad\;\Lambda}\text{Z}\,\right)\;.
	\label{eq:LL_int}
\end{align}
The calculation of these quantities proceeds in the same way of
those for single $\Lambda$~hypernuclei, starting from the
energy of the nucleus, the corresponding $\Lambda$~hypernucleus
and now the double $\Lambda$~hypernucleus.  In Table~\ref{tab:BLL},
we report the total energies for $^4$He, $^5_\Lambda$He and
$^{\;\;\,6}_{\Lambda\Lambda}$He in the second column, the single or
double hyperon separation energies in the third and the incremental
$\Lambda\Lambda$ energy in the last column. The value of $B_{\Lambda\Lambda}$
confirms the weak attractive nature of the $\Lambda\Lambda$
interaction~\cite{Hiyama:2002,Nagels:1979,Maessen:1989,Rijken:1999}.
Starting from $^4$He and adding two hyperons with
$B_\Lambda=3.22(14)$~MeV, the energy of $^{\;\;\,6}_{\Lambda\Lambda}$He
would be 1.0 to 1.5~MeV less than the actual AFDMC result.  Therefore
the $\Lambda\Lambda$ potential of Eq.~(\ref{eq:V_LL}) induces a net
attraction between hyperons, at least at this density.

Our results for  $B_{\Lambda\Lambda}$ and $\Delta B_{\Lambda\Lambda}$
are very close to the expected results for which the potential
has originally been fitted within the cluster model.  The
latest results $B_{\Lambda\Lambda}=6.91(0.16)$~MeV and $\Delta
B_{\Lambda\Lambda}=0.67(0.17)$~MeV of Ref.~\cite{Ahn:2013} suggest
a weaker attractive force between the two hyperons. A refit of
the interaction of the form proposed in Eq.~(\ref{eq:V_LL}) would be
required. It would be interesting to study other double $\Lambda$
hypernuclei within the AFDMC framework with the $\Lambda N$, 
$\Lambda NN$ and $\Lambda\Lambda$ interaction proposed. 
Some experimental data are available in the 
range $A=7 - 13$, but there are uncertainties in
the identification of the produced double $\Lambda$~hypernuclei,
reflecting in inconsistencies about the sign of the $\Lambda\Lambda$
interaction~\cite{Dover:1991,Yamamoto:1991}. An ab-initio analysis of
these systems might put some constraints on the hyperon-hyperon force,
which at present is still poorly known, and give information on its
density dependence.  Also the inclusion of the $\Lambda\Lambda N$ force
would be important.

\begin{table}[ht]
\caption[]{Comparison between $^4$He and the corresponding single
and double $\Lambda$~hypernuclei.  In the second column the total
energies are reported.  The third column shows the single
or double $\Lambda$~separation energies.  In the last column the
incremental $\Lambda\Lambda$ energy $\Delta B_{\Lambda\Lambda}$ is reported.
All the results are obtained using
the complete two- plus three-body (set~(\hyperlink{par_II}{II}))
hyperon-nucleon interaction with the addition of the $\Lambda\Lambda$
force of Eq.~(\ref{eq:V_LL}). The results are expressed in MeV.
\label{tab:BLL}}
\begin{ruledtabular}
\begin{tabular}{cccc}
System   &   {$E$}    & {$B_{\Lambda(\Lambda)}$} & $\Delta B_{\Lambda\Lambda}$ \\
\hline
		\hspace{0.7em}$^4$He            & -32.67(8)  &            ---           &          ---                \\ 
		\hspace{0.6em}$^5_\Lambda$He    & -35.89(12) &          3.22(14)        &          ---                \\ 
		$^{\;\;\,6}_{\Lambda\Lambda}$He & -40.6(3)   &          7.9(3)          &         1.5(4)              \\ 
		\midrule
		$^{\;\;\,6}_{\Lambda\Lambda}$He & {Expt.~\cite{Takahashi:2001}} & {$7.25\pm 0.19^{+0.18}_{-0.11}$} & {$1.01\pm 0.20^{+0.18}_{-0.11}$} \\
\end{tabular}
\end{ruledtabular}
\end{table}

\section{Conclusions}
\label{sec:conclusions}

We presented a detailed study of single $\Lambda$~hypernuclei
in the framework of the Quantum Monte Carlo method.  By accurately refitting
the three-body hyperon-nucleon interaction we obtain substantial
agreement with available experimental data. The present results 
confirm that the repulsion induced
by the $\Lambda NN$ force properly corrects the saturation property of
the hyperon separation energy that is strongly overestimated by
the use of a bare $\Lambda N$ interaction.

A $\Lambda\Lambda$ effective interaction has also been applied to
the study of $^{\;\;\,6}_{\Lambda\Lambda}$He. Results are in good agreement with
the available experimental data. 
This is a first step in the study of $S=-2$
$\Lambda$~hypernuclei with QMC calculations, for which there are
controversial results both from theoretical and experimental studies.

The three-body $\Lambda NN$ interaction used in this work provides a stronger
repulsion than in our previous more qualitative results.  On the
grounds of this observation, we feel confident that the application of the
$\Lambda N+\Lambda NN$ (and possibly $\Lambda\Lambda$) interaction to the
study of the homogeneous medium will lead to a stiff equation of state
for the $\Lambda$ neutron matter. This fact helps to understand how the
necessary appearance of hyperons at some value of 
the nucleon density in the inner core of a neutron star
might eventually be compatible with the observed neutron star
masses of order $2~M_\odot$~\cite{Demorest:2010,Antoniadis:2013}.
A study along this direction is in progress and encouraging results are
indeed already available.

\section{Acknowledgments}
\label{sec:acknowledgments}

This work has been partially performed at LISC, Interdisciplinary
Laboratory for Computational Science, a joint venture of the University
of Trento and Bruno Kessler Foundation.  Support and computer time
were partly made available by the AuroraScience project (funded by the
Autonomous Province of Trento and INFN), and by Los Alamos Open Supercomputing.
This research used also resources of the National Energy Research
Scientific Computing Center, which is supported by the Office of
Science of the U.S. Department of Energy under Contract No.
DE-AC02-05CH11231.
The work of S.~G. was supported by the Department of Energy Nuclear
Physics Office, by the NUCLEI SciDAC program, and by a Los Alamos LDRD
early career grant.

\bibliographystyle{apsrev4-1}

\end{document}